\documentclass[a4paper]{jpconf}

\usepackage{iopams}

\newcommand{\lanln}[1]{{\it Preprint\/} #1}
\newcommand{\SLtwor}{{\mathrm{SL}}(2,\mathbb{R})}
\newcommand{\sltwor}{{\mathfrak{sl}}(2,\mathbb{R})}
\newcommand{\BbbR}{\mathbb{R}}
\newcommand{\BbbZ}{\mathbb{Z}}
\newcommand{\BbbC}{\mathbb{C}}
\newcommand{\be}{\begin{eqnarray}}
\newcommand{\ee}{\end{eqnarray}}
\newcommand{\nn}{\nonumber}

\newcommand{\tfrac}[2]{{\textstyle \frac{#1}{#2}}}

\newcommand{\Gammared}{{\Gamma_\mathrm{red}}}

\newcommand{\Aobs}{\mathcal{A}_\mathrm{obs}}

\newcommand{\Haux}{{\mathcal{H}_\mathrm{aux}}}

\begin{document}
\title{Group averaging, 
positive definiteness 
and 
superselection sectors\footnote{Based on a talk given at the 
Fourth Meeting on Constrained Dynamics and Quantum Gravity
(Cala Gonone, Sardinia, Italy, 12--16 September 2005). 
}}

\author{Jorma Louko}

\address{School of Mathematical Sciences, 
University of Nottingham, 
Nottingham NG7 2RD, 
United Kingdom}

\ead{jorma.louko@nottingham.ac.uk\\[0.5cm] arXiv:gr-qc/0512076}

\begin{abstract}
I discuss group averaging as a method for quantising 
constrained systems whose gauge group is a noncompact Lie group. 
Focussing on three case studies, I address the convergence 
of the averaging, possible 
indefiniteness of the prospective
physical inner product and the emergence of superselection
sectors. 

\end{abstract}

\section{Introduction}

In quantisation of constrained systems, one approach to finding 
gauge invariant states is to first build an unconstrained quantum
theory and then to average states in this theory over the action of
the gauge group 
\cite{higuchi,KL,Lands-Bialo,lands-against,%
lands-wren,QORD,epistle,BC,GM1,GM2,GoMa,%
LouRov,LouMol1,LouMol2,LouMol3,Giulini-rev,Marolf-MG}. When the gauge
group is a compact Lie group with a unitary action on the
unconstrained Hilbert space, the mathematical setting is
well-understood: the averaging converges and yields a projection
operator to the Hilbert subspace of gauge invariant states. When the
gauge group is a \emph{noncompact\/} Lie group, the averaging can
in certain circumstances be interpreted in terms of
`distributional' states 
\cite{QORD,BC,GoMa,LouRov,LouMol1,LouMol2,LouMol3}, 
but at present few general results \cite{GM2} are known
as to when such circumstances can be expected to occur. When
the gauge group is not a Lie group, the situation is even more open,
although some case studies 
are known \cite{epistle,noui-perez-3d}
and a formalism relating group averaging to BRST quantisation has been
developed~\cite{Shvedov}. There are also close connections 
between group averaging, projection operator 
quantisation \cite{Klauder-proj1} 
and the master constraint programme 
\cite{Thiemann-master,ThieDitt-master1}. 

In this contribution I will discuss averaging over a noncompact Lie
group, focussing on three systems whose classical phase space is
finite dimensional
\cite{LouRov,LouMol1,LouMol2,LouMol3}. The conclusions will come in
two parts. First, if something can go wrong \cite{AIR-Murphy} in the
mathematics of group averaging, it tends indeed to do so. In
particular, in section \ref{sec:sltwor} we will discuss a system in
which the would-be physical inner product produced by the averaging
turns out to be indefinite. Second, when the mathematics of the
averaging goes wrong, there tends to be something subtle
already in the classical properties of the system. 
For example, in the system
with the indefinite would-be inner product, the classical solution
space consists of just three points. In section
\ref{sec:ash-horo} we will discuss a system in which 
isolated singular subsets in the classical solution space make it
necessary to renormalise the averaging, 
and this renormalisation
will result into superselection sectors in the quantum theory. Finding
a more general setting for such pathologies presents an 
interesting challenge for the future.

The status of group averaging as of approximately 2000 has been 
comprehensively reviewed in
\cite{Giulini-rev,Marolf-MG}. 
I shall aim at a qualitatively self-contained discussion of the
systems in
\cite{LouRov,LouMol1,LouMol2,LouMol3}.

\section{$\SLtwor$ system}
\label{sec:sltwor}

\subsection{Classical system}

We consider a system \cite{LouRov,LouMol1,MRT}
whose action reads 
\be
S=\int 
dt \, 
\bigl( 
\boldsymbol{p} \cdot \dot{\boldsymbol{u}}
+ 
\boldsymbol{\pi} \cdot \dot{\boldsymbol{v}}
-NH_1 -MH_2 -\lambda D
\bigr)
\ \ ,
\label{eq:act1}
\ee
where $\boldsymbol{u}$ and $\boldsymbol{p}$ are real vectors of
dimension~$p\ge1$, $\boldsymbol{v}$ and $\boldsymbol{\pi}$ are real
vectors of dimension $q\ge1$ and the overdot denotes differentiation
with respect to~$t$. $\boldsymbol{p}$~and $\boldsymbol{\pi}$ are
respectively the momenta conjugate to $\boldsymbol{u}$
and~$\boldsymbol{v}$ and the phase space is $\Gamma \simeq
T^*\mathbb{R}^{p+q}$.
$N$, $M$ and $\lambda$ are Lagrange multipliers
associated with the constraints
\be
H_1 &:=& \tfrac12 (\boldsymbol{p}^2-\boldsymbol{v}^2)
\ \ ,
\nn
\\
H_2 &:=& \tfrac12 (\boldsymbol{\pi}^2-\boldsymbol{u}^2)
\ \ ,
\nn
\\
D &:=& \boldsymbol{u}\cdot\boldsymbol{p}- \boldsymbol{v}\cdot\boldsymbol{\pi}
\ \ , 
\label{eq:const}
\ee
whose Poisson bracket algebra is the $\sltwor$ Lie algebra, 
\be
\{H_1\, , H_2\}&=&D
\ \ ,
\nn
\\
\{H_1\, , D\}  &=&-2H_1
\ \ ,
\nn
\\
\{H_2\, , D\}  &=&2H_2
\ \ . 
\label{eq:algebra}
\ee
The constraint algebra (\ref{eq:algebra}) can be regarded as a (much)
simplified version of that of general relativity, $H_1$ and $H_2$
being two `scalar' constraints and $D$ a single `vector'
constraint. The gauge transformations generated by the constraints are
\be
\left(
\begin{array}{c}
\boldsymbol{u}\\
\boldsymbol{p}\\
\end{array}
\right)\mapsto  g\left(
\begin{array}{c}
\boldsymbol{u}\\
\boldsymbol{p}\\
\end{array}
\right), \hspace{5ex}
\left(
\begin{array}{c}
\boldsymbol{\pi}\\
\boldsymbol{v}\\
\end{array}
\right)\mapsto  g\left(
\begin{array}{c}
\boldsymbol{\pi}\\
\boldsymbol{v}\\
\end{array}
\right),
\label{eq:gauge-transf}
\ee
where $g$ is an 
$\SLtwor$ matrix. The classical gauge group is thus $\SLtwor$. 

When $\min(p,q)>1$, the reduced phase space contains a regular part
that is a symplectic manifold of dimension $2(p+q-3)$, with a
symplectic form induced from~$\Gamma$, and certain lower-dimensional,
singular subsets. When $\min(p,q)=1$, the regular part is absent and
all that remains is singular in the sense that the symplectic form of
$\Gamma$ is pulled back to a vanishing 2-form.

$\Gamma$ admits an $\mathrm{O}(p,q)$-action that commutes with the
gauge transformations~(\ref{eq:gauge-transf}). The generators of this
action form an $\mathfrak{o}(p,q)$ algebra of functions on~$\Gamma$,
quadratic in the phase space coordinates. These functions are
observables \cite{Hen} (or perennials \cite{Kuchar-Winnipeg}) in the
sense that their Poisson brackets with the constraints
(\ref{eq:const}) vanish. It can be verified that this
$\mathfrak{o}(p,q)$ algebra contains all the information about generic
classical solutions in the sense that it separates the regular part of
the reduced phase space. It will therefore be of interest to keep
track of what happens to these $\mathfrak{o}(p,q)$ observables in the
quantum theory.

\subsection{Group averaging}

There exists a reasonably straightforward way to quantise the above
classical structure without imposing the constraints. The
unconstrained, or auxiliary, Hilbert space $\Haux$ is taken to be the
space of square integrable functions $\Psi
(\boldsymbol{u},\boldsymbol{v})$ in the inner product
\begin{equation}
(\Psi_1, \Psi_2)_\mathrm{aux}
:=
\int d^p\boldsymbol{u} \, d^q\boldsymbol{v}
\,
\overline{\Psi_1} \Psi_2
\ \ , 
\end{equation}
where the overline denotes complex conjugation. The classical
constraints (\ref{eq:const}) can be promoted into essentially
self-adjoint quantum constraints by the usual operator substitution
$\boldsymbol{p} \to -i\partial_{\boldsymbol{u}}$, $\boldsymbol{\pi}
\to -i\partial_{\boldsymbol{v}}$, and choosing in $D$ the symmetric
ordering. The commutator algebra of the quantum constraints is $i$
times the Poisson bracket algebra~(\ref{eq:algebra}), and the quantum
constraints exponentiate into a unitary representation $U$ of
$\SLtwor$ when $p+q$ is even and into a unitary representation of the
double cover of $\SLtwor$ when $p+q$ is odd. $U$~is isomorphic to the
$(p,q)$ oscillator representation of the double cover of $\SLtwor$
\cite{Howe} via the Fourier transform in~$\boldsymbol{v}$. 
Finally, the operator substitution promotes the classical
$\mathfrak{o}(p,q)$ observables into an $\mathfrak{o}(p,q)$ algebra of
densely-defined self-adjoint operators that commute with~$U$.

We now wish to implement group averaging via the map
\begin{equation} 
\eta: \phi \mapsto 
\int dg \, \phi^\dag U(g)
\ \ , 
\label{eq:eta-formal}
\end{equation}
where $dg$ is the bi-invariant Haar measure and the right-hand side is
to be interpreted in terms of its action on suitable states
in~$\Haux$. (That this map is chosen antilinear rather than linear is
just a convention.) The task is to give (\ref{eq:eta-formal}) a
meaning that makes $\eta$ into a refined algebraic quantisation
\emph{rigging map\/} \cite{epistle,GM2}: a map satisfying certain
technical postulates that enable the image of $\eta$ to be
interpreted, in the infinite-dimensional case after Cauchy completion,
as the physical Hilbert space of the constrained theory. 
The physical inner product ${(\cdot,
\cdot)}_{\mathrm{RAQ}}$ on the image of $\eta$ will then be given by
\begin{equation}
\bigl(\eta(\phi_1), \eta(\phi_2) \bigr)_{\mathrm{RAQ}} 
:= \eta(\phi_2)[\phi_1]
\ . 
\label{phys-ip}
\end{equation}

We seek to define $\eta$ via the integral 
\begin{equation}
{( \phi_1 , \phi_2 )}_{\mathrm{ga}}
:= \int dg \, {(\phi_1 , U(g)  \phi_2 )}_{\mathrm{aux}}
\ \ . 
\label{eq:ga-form}
\end{equation}
While (\ref{eq:ga-form}) is not well defined on all of~$\Haux$, 
it is possible to find a dense 
linear subspace $\Phi \subset \Haux$, called the test space, 
on which the integral in (\ref{eq:ga-form}) converges 
in absolute value. Equation 
(\ref{eq:ga-form})
then defines on $\Phi$ a Hermitian sesquilinear form 
${( \cdot , \cdot )}_{\mathrm{ga}}$  
that is invariant under $U$ in each argument. 
When (\ref{eq:eta-formal}) is interpreted as 
\begin{equation}
\eta(\phi_1)[\phi_2]
:= 
{( \phi_1 , \phi_2 )}_{\mathrm{ga}}
\ , 
\label{eq:etadef-ga}
\end{equation}
where the square brackets denote the action of $\eta(\phi_1)$ on
$\phi_2\in\Phi$, $\eta$ is then a map from $\Phi$ to its algebraic
dual, $\Phi^*$. The image of $\eta$ consists of `distributional'
states in the sense that $\Phi^*$ is not contained in the Hilbert dual
of~$\Haux$. It follows that the image of $\eta$ consists of states
that are gauge invariant, in the sense that for every $\phi \in \Phi$
and $g$ in the gauge group, a state in the image of $\eta$ has the
same action on $U(g)\phi$ and~$\phi$. It is further possible to choose
$\Phi$ to be invariant under the quantum $\mathfrak{o}(p,q)$
observables and to satisfy certain technical conditions introduced in
\cite{GM2}. This means that we will obtain a quantum theory if just
two more conditions can be shown to hold: The image of
$\eta$ needs to be nontrivial and the sesquilinear form
(\ref{phys-ip}) on this image needs to be positive definite.

The outcome depends sensitively on $p$ and~$q$: 
\begin{itemize}
\item
From the viewpoint of the classical theory, the most interesting case
is $\min(p,q)>1$. For $p+q$ even, we do recover a quantum theory, and
this theory carries a nontrivial representation of the quantum
$\mathfrak{o}(p,q)$ observables. For $p+q$ odd, by contrast, the image
of $\eta$ is trivial and we obtain no quantum theory. Quantisation for
$p+q$ odd would seem to require some modification that allows the test
states to have
\emph{noninteger\/} 
angular momenta in $\boldsymbol{u}$ or~$\boldsymbol{v}$.
\item
From the viewpoint of the classical theory, one would expect something
pathological to happen when $\min(p,q)=1$. This turns out to be the 
case. We obtain either no quantum theory or a quantum theory with a
one-dimensional Hilbert space and a trivial representation of the 
quantum $\mathfrak{o}(p,q)$ observables. 
\end{itemize}

For $(p,q)=(1,1)$, the sesquilinear forms (\ref{phys-ip}) and
(\ref{eq:ga-form}) are nonvanishing but have
\emph{indefinite\/}
signature.  The image of $\eta$ is nontrivial and two-dimensional, but
the would-be inner product (\ref{phys-ip}) on this image is
indefinite. This is the first example known to the author in which
group averaging fails by producing an indefinite would-be inner
product.

\section{Triangular $\SLtwor$ system}

We next consider a system \cite{LouMol2} obtained from the $\SLtwor$
system of section \ref{sec:sltwor} by dropping $H_1$ from the
action~(\ref{eq:act1}). The classical gauge group is then 
the connected component of the lower
triangular subgroup of $\SLtwor$. 

The interest in this system arises from the observation that since the
gauge group is not unimodular, the left and right Haar measures do
not coincide, and neither of these measures is invariant under the
group inverse. In order for the group averaging to produce in
(\ref{eq:ga-form}) a 
Hermitian sesquilinear form, the integration measure needs
to be invariant under the group inverse, and a measure with this property
is the geometric average $d_0g$ of the left and right invariant Haar
measures. As noted in \cite{GM2}, adopting
$d_0g$ in (\ref{eq:ga-form}) means that the physical states will
\emph{not\/} 
be invariant under the gauge group action. 
Instead, the physical
states will satisfy the constraints in a sense that in certain classes
of systems has been shown to be equivalent to geometric quantisation
in the reduced phase space \cite{Tuyn,DEGT,DEGST}.

The triangular $\SLtwor$ system is classically singular for
$(p,q)=(1,1)$ in the sense that the pull-back of the phase space
symplectic form to the reduced phase space vanishes. For
$(p,q)\ne(1,1)$ the reduced phase space has still singularities, but
these singularities form a set of measure zero and the rest is a
symplectic manifold, symplectomorphic to $T^*(S^{p-1} \times
S^{q-1})$. In this sense the system is classically regular for
$(p,q)\ne(1,1)$.

The system possesses the same $\mathfrak{o}(p,q)$ algebra of classical
observables as the $\SLtwor$ system of section~\ref{sec:sltwor}. For
$(p,q)=(1,1)$ these observables vanish on all of reduced phase space,
but for $(p,q)\ne(1,1)$ they separate the regular part of the reduced
phase space up to a certain measure zero subset and an overall twofold
degeneracy. As in the $\SLtwor$ system, it will therefore be of
interest to keep track of what happens to these observables on
quantisation.

Group averaging with the measure $d_0g$ in (\ref{eq:ga-form}) can now
be completed for all $(p,q)$ and yields in each case a quantum theory
that carries a maximally degenerate principal unitary series
representation of the quantum $\mathfrak{o}(p,q)$
observables~\cite{Vil}. The representation is nontrivial iff $(p,q)
\ne (1,1)$, that is, precisely when the classical system is regular in
the above sense.

\section{Ashtekar-Horowitz-Boulware system}
\label{sec:ash-horo}

Our third system \cite{LouMol3} was introduced by Boulware
\cite{Boul} as a simplifed version of the Ashtekar-Horowitz model
\cite{AH}, as a venue for studying tunnelling phenomena in
constrained quantisation
\cite{Boul,AH,Gotay,Tate, Ash2}. 
While our results on tunnelling coincide with those in \cite{Boul},
the new issue of interest for us is in the sense of convergence of the
group averaging and in the resulting superselection sectors in the
quantum observable algebra.

The configuration space of the classical system is $S^1 \times S^1 =
\bigl\{(x,y)\bigr\}$, where $x$ and $y$ are understood periodic with
period~$2\pi$. The action reads
\be
S=\int\! dt\, \bigl(  p_x \dot{x} + p_y \dot{y} 
- \lambda C \bigr)
\ ,
\label{eq:hamilt}
\ee
where 
$\lambda$ is a Lagrange multiplier associated with 
the constraint
\be
C := p_x^2 - R(y) 
% \ ,
\label{eq:class-constraint}
\ee
and the potential 
$R: S^1 \to \BbbR$ is a smooth function that is positive at least
somewhere and satisfies certain
genericity conditions. In particular we assume $R$ to have finitely
many zeroes and stationary points, all stationary points to have
finite order and no zero to be a stationary point. 
It follows that the reduced phase space
$\Gammared$ is a two-dimensional 
symplectic manifold with certain singular points and
finite symplectic volume. The singularities are located at the
stationary points of $R$, and they are topological in nature, arising
from the periodicity of~$x$. 

We choose the auxiliary Hilbert space 
$\Haux$ to be the space of square integrable functions of the
configuration variables in the inner product 
\begin{equation}
\left( \phi_1 ,\phi_2 \right)_{\mathrm{aux}}
:=
\int \! dx \, dy\, \overline{\phi_1
(x,y)}\phi_2 (x,y)
\ . 
\end{equation}
The classical constraint (\ref{eq:class-constraint}) is promoted into
the essentially self-adjoint quantum constraint
\be
\hat{C}:= - \frac{\partial^2}{\partial x^2} - \hat{R}
\ , 
\ee
which exponentiates into the one-parameter family of unitary operators 
\be
U(t):=e^{-it\hat{C}}
\ , 
\quad
t\in\BbbR
\ . 
\ee
The gauge group of the quantum theory is thus~$\BbbR$. 

For the test space $\Phi \subset \Haux$, in which formula
(\ref{eq:eta-formal}) is to be given a meaning, we adopt the space of
functions of the form $f(x,y) =
\sum_{m\in\BbbZ} e^{imx} f_m(y)$, where each $f_m: S^1 \to \BbbC$ is
smooth and only finitely many $f_m$ are different from zero for
each~$f$. Proceeding for the moment 
formally, (\ref{eq:eta-formal}) then yields 
\be
\bigl(\eta (f) \bigr) (x,y)
=  
2\pi \sum_{mj} 
\frac{e^{-imx} \overline{f_m (y)} }
{\left| R'(y_{|m|j}) \right|}\, 
\delta(y,y_{|m|j})
\  ,
\label{eq:eta-ftilde1}
\ee
where $y_{|m| j}$ are the solutions to the equation 
\begin{equation}
m^2 = R(y) 
\label{eq:m2=R}
\end{equation}
and the delta-distribution on the right-hand side of
(\ref{eq:eta-ftilde1}) is that on~$S^1$. The task is now to examine
whether equation (\ref{eq:eta-ftilde1}) in fact defines a rigging
map. A~necessary condition clearly is that (\ref{eq:m2=R}) has
solutions for some $m\in\BbbZ$. 

Suppose first that no $y_{|m| j}$ is
a stationary point of~$R$. Most of the rigging map properties can then
be immediately verified from~(\ref{eq:eta-ftilde1}). In
particular, the image of $\eta$ is a nontrivial finite-dimensional 
vector space, 
states in the image of $\eta$ 
are gauge invariant in the sense that 
$\eta(f)[U(t)g] = \eta(f)[g]$ for all $f,g\in\Phi$ and
$t\in\BbbR$, and the sesquilinear form defined on the image of $\eta$
by (\ref{phys-ip}) is positive definite.

The rigging map property that is not immediate from
(\ref{eq:eta-ftilde1}) is whether $\eta$ induces a representation of
the refined algebraic quantisation observable algebra $\Aobs$ on the
physical Hilbert space. The operators in $\Aobs$ are densely defined
on $\Haux$ and required to be gauge invariant in the sense that they
commute with $U(t)$ for all~$t$, but they are also required to satisfy
certain technical conditions, in particular that their domain includes
$\Phi$ and they map $\Phi$ to itself. The condition for $\eta$ to
induce a representation of $\Aobs$ on the physical Hilbert space reads
\be
\eta(A \phi_1) [\phi_2]
= 
\eta(\phi_1) [A^\dag \phi_2]
\label{eq:eta-intertwining-matrix}
\ee
for all $\phi_1, \phi_2 \in \Phi$ and $A \in \Aobs$. As the
definition of $\Aobs$ is implicit rather than constructive,
verifying (\ref{eq:eta-intertwining-matrix}) involves some 
subtlety but can be accomplished 
by considering matrix elements of the form 
${\bigl(A f, U(t) g\bigr)}_{\mathrm{aux}}$. Hence $\eta$ is a rigging
map. It can be verified that the representation of $\Aobs$
on the resulting physical Hilbert space is irreducible, and in fact
transitive. 

Suppose then that some solutions to (\ref{eq:m2=R}) are stationary 
points of~$R$. The corresponding terms in (\ref{eq:eta-ftilde1}) have
a zero in the denominator and are hence ill defined. However, under
appropriate technical conditions on the stationary point structure
of~$R$, these formally divergent terms can be replaced by finite ones
that involve fractional powers of higher derivatives of $R$ in the
denominator. This replacement can be understood as a renormalisation
of the averaging by an infinite multiplicative constant whose
`magnitude' depends on the order of the stationary point. We find that
including in (\ref{eq:eta-ftilde1}) only the non-stationary solutions
to (\ref{eq:m2=R}) yields one rigging map, but there are also others, 
given by the renormalised terms in which the solutions
to (\ref{eq:m2=R}) are respectively (i) maxima of a given order, (ii)
minima of a given order, and (iii) points of inflexion of a given
order. Each of these rigging maps leads to a physical Hilbert space
that carries a transitive representation of~$\Aobs$. 

As the images of any two of these rigging maps have trivial
intersection in~$\Phi^*$, we can regard the direct sum of the
individual Hilbert spaces as the `total' physical Hilbert space
$\mathcal{H}_\mathrm{RAQ}^{\mathrm{tot}}$, 
with a representation of $\Aobs$ that decomposes into the 
representations on the summands. The
summands thus constitute superselection sectors
in~$\mathcal{H}_\mathrm{RAQ}^{\mathrm{tot}}$.

The emergence of superselection sectors is related to the
singularities in~$\Gammared$. The coordinate $x$ is periodic, and the
conjugate momentum $p_x$ has become quantised in integer values. For
generic potentials, these integer values entirely miss the
singular, measure zero subsets of~$\Gammared$, and in this case the
quantum theory has no superselection sectors. Superselection sectors
appear precisely for those potentials for which some of the
quantised values of $p_x$ hit a singular subset
of~$\Gammared$.

Finally, consider the semiclassical limit. Let
$\mathcal{H}_\mathrm{RAQ}^{1}$ denote the Hilbert space that comes
from the nonstationary solutions to~(\ref{eq:m2=R}). When $\hbar$ is
restored, the dimension of $\mathcal{H}_\mathrm{RAQ}^{1}$ asymptotes
in the $\hbar\to0$ limit to $(2\pi\hbar)^{-1}$ times the volume
of~$\Gammared$, while the dimension of the orthogonal complement of
$\mathcal{H}_\mathrm{RAQ}^{1}$ in
$\mathcal{H}_\mathrm{RAQ}^{\mathrm{tot}}$ remains bounded. In this
sense, the semiclassical limit in
$\mathcal{H}_\mathrm{RAQ}^{\mathrm{tot}}$ comes entirely from the
superselection sector $\mathcal{H}_\mathrm{RAQ}^{1}$. Note that the
semiclassical limit is as might have been expected on comparison with
geometric quantisation on compact phase spaces
\cite{woodhouse-book,sniatycki}.

\section{Discussion and outlook}
\label{sec:discussion}

Although group averaging has proved a viable method for quantising
certain systems whose gauge group is a noncompact Lie group, 
general principles of when and how group averaging might be expected
to work remain largely open. We conclude by discussing some of these
open issues in the light of our systems.

A central ingredient in the quantisation is the choice of the test
space $\Phi \subset \Haux$. This space has a dual role. On the one
hand, the test space is a technical necessity, since the integral over
a noncompact gauge group is not expected to converge on all
of~$\Haux$. On the other hand, the test space has a deep physical
significance in that it determines the observables of the quantum
theory. Refined algebraic quantisation does not as such require a
single classical or quantum observable to be explicitly constructed,
but if some classical observables of interest are known, then it is
the choice of the test space that determines whether corresponding
observables will exist in the quantum theory. This issue arises
prominently in our $\SLtwor$ system, where considerable tuning of the
test space is required to guarantee that a quantisation of the
classical $\mathfrak{o}(p,q)$ observable algebra is included in the
observable algebra of the quantum theory. General conditions under
which technically viable and physically appropriate test spaces exist
remain an open problem.

The $\SLtwor$ system with $(p,q)=(1,1)$ provides an example in which
the group averaging sesquilinear form (\ref{eq:ga-form}) is
indefinite. Any general theorems one might wish to prove on group
averaging therefore do need to take this eventuality into account. The
Giulini-Marolf uniqueness theorem \cite{GM2} is a perspicacious 
case in point,
implying that when group averaging converges sufficiently strongly to
an indefinite sesquilinear form, the system admits no rigging maps. It
would be of interest to understand just how special the $\SLtwor$
system with $(p,q)=(1,1)$ is. The system is classically quite
singular, with a reduced phase space that contains just three points,
non-Hausdorff close to each other. Can an indefinite would-be inner
product arise in classically well-behaved systems?

The Ashtekar-Horowitz-Boulware system displays a situation in which
group averaging does not converge in absolute value, and with some
potentials not even in a conditional sense. Nevertheless, the
averaging can be consistently renormalised, with the consequence that
the representation of the physical observable algebra decomposes into
superselection sectors. This phenomenon has been observed previously
\cite{GoMa}, but what is striking in the Ashtekar-Horowitz-Boulware
system is that the superselection sectors appear precisely when some
of the discrete eigenvalues of a certain operator take values at which
the corresponding classical observable hits a singularity in the
reduced phase space. Is there a more general connection between
classical singularities and quantum superselection sectors?

From the gravitational viewpoint, systems whose gauge group is a Lie
group tend to arise in symmetry reductions of gravity, as is the case
with spatially homogeneous cosmologies, or in systems that have been
constructed by hand to mimic certain aspects of gravity, as is the
case with all the systems discussed in this contribution. In gravity
proper, however, the gauge group is infinite dimensional, and the
Poisson bracket algebra of the constraints closes not by structure
constants but by structure functions. While group averaging with
nonunimodular Lie groups may give some insight into structure
functions \cite{GM2}, and while a formalism that ties group averaging
to BRST techniques has been developed~\cite{Shvedov}, an extension of
group averaging to systems with structure functions remains yet to be
developed to a level that would allow a precise discussion of
convergence properties and the observables in the ensuing quantum
theory.

\ack

I thank Carlo Rovelli and Alberto Molgado for collaboration and the
QG05 organisers for the invitation to present this work. During the
meeting I benefited from discussions on related topics with a number
of participants, including Steve Carlip, Marco Cavagli\`a, Theodosios
Christodoulakis and John Klauder.

\end{document}